\magnification=1200
\baselineskip=17truept
\input epsf

%preprint or not
\def\preprint{Y}
%draft or not
\def\draftversion{N}
\def\cap{\hsize=4.6in}

\if \draftversion Y


\fi

% Figure
\def\figure#1#2#3{\if \preprint Y \midinsert \epsfxsize=#3truein
\centerline{\epsffile{figure_#1eps}} \halign{##\hfill\quad
&\vtop{\parindent=0pt \hsize=4.6in \strut## \strut}\cr {\bf Figure
#1}&#2 \cr} \endinsert \fi}

\def\captionone{\cap
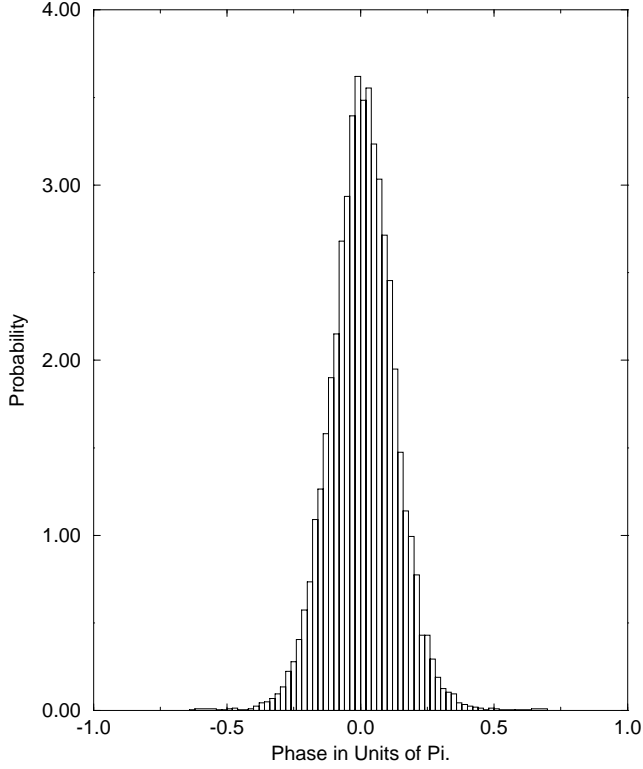
Phase distribution along a generic orbit carrying zero topological charge.
The horizontal axis is in units of $\pi$.  The phase is measured relative to
the Landau gauge phase and, within errors, the average cancels the Landau
phase leaving an almost real answer. The histogram contains 10,000
points.}

\def\captiontwo{\cap
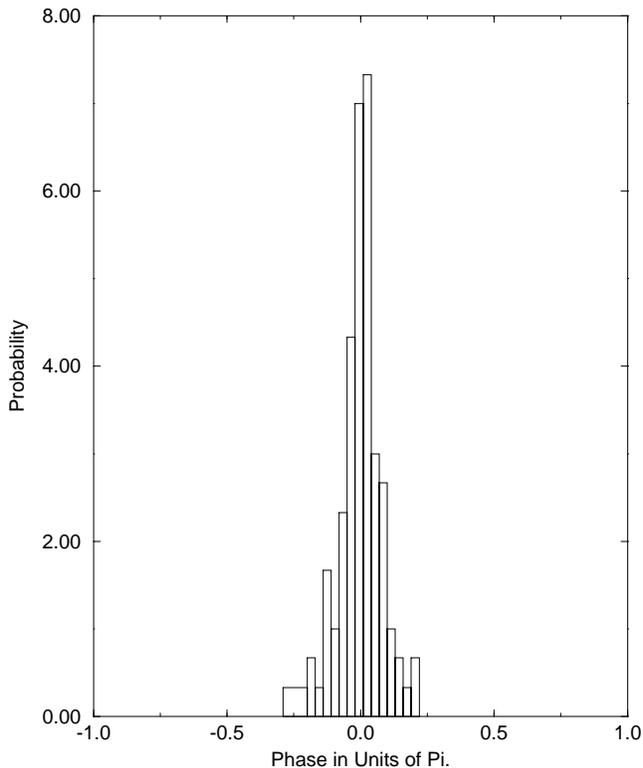
Phase distribution along a generic orbit carrying unit topological charge.
The horizontal axis is in units of $\pi$. The phase is measured relative to
the Landau gauge phase and, within errors, the average cancels the Landau
phase leaving an almost real answer. The histogram contains 100 points.}

\line{\hfill UW/PT--96--18}
\line{\hfill DOE/ER/40561-285-INT96-00-145} 
\line{\hfill RU--96--84}
\vskip 2truecm
\centerline{\bf Massless Composite Fermions in Two Dimensions and the Overlap.}

\vskip 1truecm
\centerline{Rajamani Narayanan$^{1}$ and Herbert Neuberger$^{2}$}
\vskip 0.5truecm
\centerline{${}^1$ \it Institute for Nuclear Theory, Box 351550}
\centerline{\it University of Washington, Seattle, WA 98195-1550.}
\vskip .1in
\centerline{${}^2$\it Department of Physics and Astronomy}
\centerline{\it Rutgers University, Piscataway, NJ 08855-0849. }
\vfill
\centerline{\bf Abstract}
\vskip 0.75truecm
There exist chiral gauge models in two dimensions that have massless
composite fermions. Two examples are presented and it is suggested that they
be accepted as benchmark test-cases for generic proposals of
non-perturbatively regulating chiral gauge theories in any dimension.
We apply the overlap to the simpler of the two benchmarks and present
the results of a numerical simulation of modest size. 

\vfill
\eject

Chiral symmetries can ensure the masslessness of fermions; thus, chiral
gauge theories might produce massless composite fermions. A major
objective of lattice formulations of chiral gauge theories is to provide
a tool to study this non-perturbative phenomenon numerically. Given a 
proposal, before jumping to costly simulations in four dimensions,
one needs to establish confidence in the technique and a natural theoretical
laboratory is offered by the two dimensional world. A proposal that
fails in two dimensions is unlikely to be useful in four.

The possibility of composite massless fermions was intensely studied
in the late seventies and early eighties [1]. 
The major ideas in this context were
the maximal attractive channel (MAC) hypothesis, complementarity between 
the Higgs and the confining phase descriptions and fulfillment of 't Hooft's
consistency conditions [2]. In search of a 
dynamically controlled environment
where these ideas could be tested and applied the two dimensional laboratory
was visited already at that time, 
but the results were somewhat disappointing [3].

Today, when we again turn to two dimensions, we must look for models that, 
not withstanding the earlier work, nevertheless provide interesting 
test-cases for numerical techniques.  Our first example has already been
put to use by us before [4], but its major dynamical
features that single it out for our purposes were not described. 

The model (``11112'') contains four left-handed Weyl fermions (LW) carrying
a $U(1)$ charge 1 and one right-handed Weyl fermion (RW) of charge 2. Anomalies
cancel for this $U(1)$ and it can be gauged. The model also has an anomalous
global $U(4)$ symmetry. Consider first the $SU(4)$ ($Spin (6)$) component:
The constituent anomaly would match with composites making up a six dimensional
real representation (second rank antisymmetric in $SU(4)$ language) by 
left-handed Majorana-Weyl (LMW) fermions. If there are no other massless
particles gravitational anomalies also match. The $U(1)$ component of the
global $U(4)$ is broken explicitly by instantons in the gauged $U(1)$, so
there is no matching condition associated with it. These kinematic 
considerations are supported by the exact solution of the model:

The action of the 11112 model in Euclidean space is:
$$
S={1\over 4e_0^2} \int d^2 x F^2_{\mu\nu}
-\sum_{k=1}^4\int d^2 x \bar\chi_k \sigma_\mu 
(\partial_\mu+iA_\mu )\chi_k
-\int d^2 x \bar\psi \sigma^*_\mu (\partial_\mu+2iA_\mu )\psi ,$$
where $\sigma_1=1$ and $\sigma_2=i$.
The candidate interpolating fields for the composites are:
$$
\eta_{ij} = \chi_i \chi_j \bar\psi , ~~~
\bar\eta_{ij}=\bar\chi_i \bar\chi_j \psi .$$
As long as 
the gauge field $A_\mu \equiv \partial_\mu \Phi +
\epsilon_{\mu\nu} \partial_\nu \phi$ carries zero topological charge
the path integral can be done by changing
variables: $\chi_i = \chi_i^0 e^{-i\Phi -\phi}$, 
$\bar\chi_i = \bar\chi_i^0 e^{i\Phi + \phi}$,
$\psi = \psi^0 e^{-2i\Phi+2\phi}$, 
$\bar\psi = \bar\psi^0 e^{2i\Phi-2\phi}$.
All fields with a zero superscript have a free field action. The 
transformation has a Jacobian which together with the original pure
gauge term gives $S_g = {1\over {2e_0^2}}\int d^2 x (\partial^2 \phi )^2
-{2\over\pi} \int d^2 x
\phi \partial^2 \phi$. Thus the path integral is quadratic
at each stage [5]. The disappearance of $\Phi$ reflects anomaly cancelation for
the gauged $U(1)$. 

We first evaluate the fermion number conserving
two point function of the composites:
$$
<\eta_{ij} (x) \bar \eta_{kl} (y)> = ( \delta_{ik}\delta_{jl}-
\delta_{il}\delta_{jk}) ~
{1\over{(2\pi)^3}} ~ {1\over {(\sigma\cdot z)^2}} ~{1\over {\sigma^* \cdot z}}
~e^{8<[\phi (x) - \phi (y)]^2>} ~~,$$
where $z=x-y$. Using
$$ <\phi (x) \phi (y)>-<\phi^2 (0)> =-{1\over 16}
\left[\int_0^\infty {{dt}\over t}
e^{-{1\over t} - {{t e_0^2 z^2} \over \pi}} +2\gamma +
\log {{e_0^2 z^2}\over \pi}\right ]
\equiv -{1\over{16}} G(z^2)$$ 
we get
$$e^{8<[\phi (x) - \phi (y)]^2>}= e^{G(z^2)}.
$$
As $z^2 \to\infty$, we obtain:
$$
<\eta_{ij} (x) \bar \eta_{kl} (y)> ~ \sim ~ ( \delta_{ik}\delta_{jl}-
\delta_{il}\delta_{jk})~ 
{e_0^2 \over {4\pi^3}}~
e^{2\gamma}~ {1\over{2\pi\sigma\cdot z}}~[1 +{\cal{O}} (e^{-{{2e_0 |z|}\over
{\sqrt{\pi}} } } ) ].
$$
Normalizing the interpolating fields $\eta$ we see that we have a massless
pole corresponding to a left-mover. Note that the corrections are exponential,
so no couplings to other massless particles are evident (absence of cuts). 

To ascertain that we have indeed LMW particles and not LW particles we
need to also compute the $<\eta \eta >$ and $< \bar\eta \bar\eta >$ correlation
functions. These propagators don't vanish since fermion 
number conservation is broken
by instanton effects. More precisely, single instantons allow violations
by two units, so indeed the particle and anti-particle created by an $\eta$
can mix. One cannot calculate the new propagators by the above method
directly since nontrivial topology is needed. However, using clustering, the
infinite volume computation (which is all we are doing) can be replaced
by a calculation at trivial topology. We start from
$<\eta_{12} (x) \eta_{34} (x+\Delta ) \bar\eta_{12} (y) 
\bar\eta_{34} (y+\Delta ) >$. 
The computation proceeds as before and 
an exact expression is obtained:
$$ <\eta_{12} (x) \eta_{34} (x+\Delta ) \bar\eta_{12} (y) 
\bar\eta_{34} (y+\Delta ) > = $$
$$
{1\over{(2\pi)^6}}~{1\over{(\sigma\cdot z)^4}}~ {{(\sigma^* \cdot \Delta)^2}
\over {(\sigma^* \cdot z )^2 [(\sigma^* \cdot z )^2 -
(\sigma^* \cdot \Delta )^2 ]}} ~ e^{-2G (\Delta^2 ) +
G(|z+\Delta|^2) + G(|z-\Delta|^2) +2 G (z^2 )}.$$
We take now the separation $z=x-y$ to infinity and, asserting that clustering
holds obtain:
$$
<\eta_{12} (x) \eta_{34} (x+\Delta )>< \bar\eta_{12} (y) 
\bar\eta_{34} (y+\Delta )>= {1\over{(2\pi)^6}}~ e^{-2 G (\Delta^2 )}~
e^{8\gamma} ~({{e_0^2}\over\pi})^4 ~(\sigma^*\cdot \Delta )^2 .$$
Using translational invariance, Lorentz invariance, and charge conjugation,
we end up having to take a square root obtaining:
$$
<\eta_{12} (0) \eta_{34} (\Delta )>=<\bar\eta_{12} (0) \bar\eta_{34} (\Delta )>
~\sim ~{1\over{(2\pi)^3}}~ e^{- G (\Delta^2 )}~
e^{4\gamma} ~({{e_0^2}\over\pi})^2 ~(\sigma^*\cdot \Delta ).$$
The undetermined overall signs for all index combinations can be absorbed
in three of the six $\eta$-fields and we shall assume this has been done.

From the above equation we extract 
both the asymptotics as $|\Delta|\to\infty$ and as  
$|\Delta|\to 0$. Going to large separation we get
$$
<\eta_{ij} (x) \eta_{kl} (y)>
=<\bar\eta_{ij} (x) \bar\eta_{kl} (y)> ~\sim ~ -\epsilon_{ijkl}
{{e_0^2}\over{4\pi^3}} ~e^{2\gamma} ~
{1\over{2\pi\sigma\cdot z}}~[1 +{\cal{O}} (e^{-{{2e_0 |z|}\over
{\sqrt{\pi}} } } ) ].$$
The ${{e_0^2} \over{4\pi^3}} e^{2\gamma}$ prefactor is the same as in the
$< \eta\bar\eta >$ correlation function, so the same normalization makes
the propagators canonical. Going to short distances we extract the
expectation value of the leading operator:
$$
<\eta_{ij} (0) (\sigma\cdot\partial) \eta_{kl} (0)>=
\epsilon_{ijkl} ~
{{e_0^4 e^{4\gamma}}\over{4\pi^5}}.$$
Defining the dimensionless 't Hooft vertex $V={{\pi^2}\over
{e_0^4}} \chi_1 \chi_2 \chi_3 \chi_4 \psi (\sigma\cdot\partial )\psi$
we obtain
$$
<V>={{e^{4\gamma}}\over {4\pi^3}}\approx 0.081$$

The 't Hooft vertex has the form of a term in the kinetic energy for 
composite massless left-movers in a low energy effective Lagrangian.
The above expectation value arises from a single instanton background
plus small fluctuations around it.  Other terms in the kinetic energy
get an expectation value in zero topology. One cannot argue convincingly
(in lieu of the exact solution) that the required ``exponentiation''
occurs since there is no justifiable ``dilute gas approximation''.
However, if we adapt ``complementarity'' [2] to two dimensions we could say
that we have an effective scalar field made out of, say, 
$\chi_1 \bar\psi$ and its conjugate. This scalar field behaves as if it
has a negative mass, but, in two dimensions, unlike in four, it does
not condense [6]. 
This composite ``Higgs'' field in two dimensions affects the $U(1)$ gauge
dynamics in a different way from four. Instead of making the forces
short range it preserves confinement via vortex (instanton) formation.
What changes as a result of the sign of the mass 
is only the size of the string tension. Now confinement is a semiclassical
effect and the ``dilute gas'' picture can be made to hold. Within the dilute
gas picture one can argue that the 't Hooft vertex exponentiates 
in the usual way. Of course, this is only a description which, while not
really applicable, is hoped to lead to the correct conclusions
regarding the low mass spectrum. The exponentiation leads us to an effective 
Lagrangian describing LMW particles. 

Since we are in the fortunate position of having exact results we
do not need to appeal to complementarity to argue that indeed only a sextet
of LMW particles is massless. We define the linear combinations
$$
\rho^\pm_{ij} ={{\pi^{3\over 2} e^{-\gamma}}\over e_0}~
[\eta_{ij} \pm {1\over 2}\epsilon_{ijkl} \bar\eta_{kl}].$$
While
$$
<\rho^-_{ij} (x) \rho^-_{kl} (y) > ~\sim ~ -\epsilon_{ijkl} 
{1\over{2\pi\sigma\cdot z}},$$
correlators involving $\rho^+$ have no massless poles, and no cuts extending
to the origin. In short the massless spectrum consists of six LMW fermions.

Clearly, a numerical method could establish its credentials by reproducing
the above gauge invariant spectrum.  It is obvious that a scheme that does not
``quite work'' in the presence of nonzero topology would fail completely.
We proceed now to a non-Abelian model where the issue of topology does not
arise, and where a scheme could be tested even if it is known to mishandle
topology. At present we have no exact results about the non-Abelian model,
although we would not exclude the possibility that the spectrum of massless
composites could be established by analytical means also in this case.

The gauge group is $SU(2)$ and matter consists of two LW doublets 
$\chi_{i\alpha}$ and one RMW triplet $\psi_A$. $\alpha,\beta =1,2$, 
$a=1,2,3$ refer to flavor (not gauged) and $i,j,k=1,2$, $A=1,2,3$ refer to color
(gauged). The flavor symmetry group is $U(2)$ and there are no instantons.

Using $\tau$ to denote Pauli matrices we guess that the
massless composites are given by (upper or lower positions
of indices are meaningless) :
$$
\eta_a = \tau_a^{\alpha\beta} \tau_A^{ij}\bar \chi_{i\alpha} 
\chi_{j\beta}\psi_A,~~~~
\rho = \tau_A^{ij} \epsilon^{jk}\epsilon^{\alpha\beta}
\chi_{i\alpha}\chi_{k\beta}\psi_A ,~~~~
\bar\rho = \tau_A^{ij} \epsilon^{jk}\epsilon^{\alpha\beta}
\bar\chi_{i\alpha}\bar\chi_{k\beta}\psi_A $$
The $\eta$'s are a flavor $SU(2)$ triplet made out of LMW particles
and the $\rho$ is a flavor $SU(2)$ singlet LW particle. The $\eta$'s
carry no flavor $U(1)$ charge but $\rho$ is doubly charged.
It is simple to check that the flavor $U(1)$, $SU(2)$ and the
gravitational anomalies match between the constituents and the composites.

The Abelian and non-Abelian models can be connected by adding a Higgs color 
triplet to the non-Abelian one. The absence of instantons makes it plausible
that things look more like in four dimensions. Thus, we expect two of
the three gauge bosons to become massive leaving us with a confining $U(1)$.
Switching some particles with anti-particles we see that the $\chi$ fields
here play the role of the $\chi$ fields in the Abelian case and the
RW $\psi$ there is made up of $\psi_1\pm i\psi_2$ here. 
So, on the basis of the Abelian model we expect six $LMW$ particles
and, in addition, the left over massless neutral RMW $\psi_3$. One
can imagine that a Majorana mass-term connecting $\psi_3$ to one of the
six left-movers gets generated and one has the same number of massless
particles (five LMW) as in the analysis above. 

Of course, we cannot be sure that we have correctly identified the massless
spectrum in the non-Abelian case. But it is plausible that
it is just as simple as we have guessed. So, if a simulation reproduced
this spectrum one could feel encouraged to believe in the technique.

There is one issue that spoils the usefulness of two dimensions to some
extent [7]. Four fermi interactions can be marginal or even
marginally relevant in two dimensions. Typically, the symmetries of
gauge theories do not exclude these terms, and a generic regularization
will generate them. Their presence in the continuum limit can have
dramatic effects. To usefully test a procedure on two dimensional
benchmarks one would need to do some tuning to minimize the
effective couplings of all possible four fermi terms. For example,
in the Abelian model the following (Thirring) term is exactly marginal and
allowed by all symmetries:
$$
\int d^2 x \sum_k [ \bar\chi_k \chi_k ] \bar\psi \psi .$$
Its inclusion does not spoil the solubility of the model, but
does do away with the simple pole structure in the infrared, replacing
it by cuts. One no longer has particles and an $S$-matrix in the usual
sense. This is the single four fermi term possible so tuning it away
(it is only marginal - not relevant) is a nuisance one can hope to live
with.

We now turn to applying the overlap prescription to the Abelian model.
When facing a numerical simulation one needs to formulate the theory
in some finite volume, typically a torus (since otherwise one has no remnant
of translational invariance). With massless particles around 
this can be delicate but the issue has been settled for our case in our 
previous work [4]. It turns out that a good definition on a torus
consists of making $\psi$ periodic in both directions (PP) and the
four $\chi$ fields obey PP, AP, AA, PA boundary
conditions respectively (A=anti-periodic). (Note that
the global $SU(4)$ is broken by the boundary conditions although a cyclic
four element symmetry coupled to rotations is preserved. Incidentally,
this also singles out only one of the $\chi$'s (PP)
as a possible partner
of $\psi$ in the complementary, effective Higgs field picture).

Having settled the issue of boundary conditions one is still faced with
the fact that one would need a finite 
large box to observe the power decay of the
composites' correlation functions. 
Since the composites are not pointlike, one also 
needs to keep a few lattice points ``inside'' them. This is a typical numerical
problem faced whenever one does Monte Carlo simulations. In our case,
in view of the fact that the algorithm essentially involves the full
computation of fermionic determinants, the resource requirements
are comparable to 
those of a modern large scale numerical Monte Carlo computation 
in four dimensions. 
As a first step it makes sense to devise a smaller project, which can be
still done on a single workstation in a reasonable amount of time. 
This was something we have been preparing for when we did a small
scale simulation for the vector--like Schwinger model [7,8]. The objective is
to measure $<V>$, the 't Hooft vertex expectation value in small to moderate
physical volumes. Although finite size effects are sizable, they are
computable and seeing that one gets reasonable numbers is a strong indications
that things would work out correctly in a large scale simulation also.
On the other hand, if $<V>$ does not come out right, it is very unlikely that
the correct particle spectrum will be obtained in a large scale simulation.
Moreover, the smaller project sets the correct scale, parameters 
and helps determining how to tune away the Thirring term.

Since the Thirring term will be also a nuisance in vector-like simulations,
and the latter can be simulated within significantly
less computer time, because one does not need to do numerical
gauge averaging along orbits as gauge invariance is exact, we learn how
to tune its coefficient in the 
vector--like case first. Of course, there is no guarantee
that this will really apply to the chiral case. However, there does exist a
vector--like theory that is quite close to our chiral model. It is 
the four flavor Schwinger model, defined by:
$$
S_{v}={1\over 4e_0^2} \int d^2 x F^2_{\mu\nu}
-\sum_{k=1}^4\int d^2 x \bar\chi_k \sigma_\mu 
(\partial_\mu+iA_\mu )\chi_k
-\sum_{k=1}^4 \int d^2 x \bar\psi_k \sigma^*_\mu (\partial_\mu+iA_\mu )\psi_k 
$$
At zero topology the result of integrating out the fermions on a torus is the
the same as in the 11112 model. The vector--like model also has a
't Hooft vertex $V_v$ which gets an expectation value in the single instanton
sector.
$$
<V_v> \equiv {{\pi^2}\over{e_0^4}} <\prod_k (\bar\psi_k \chi_k )> = 
{{e^{4\gamma}}\over{(2\pi )^4}} \approx .00645$$

Using the same regularization scheme for both
models one expects that the Thirring term here ($\sum_k (\bar \chi_k \chi_k )
\sum_k (\bar\psi_k \psi_k )$) will 
have an induced coefficient of similar size to
the one in the chiral version. 
In the vector-like case, depending on the sign of the 
Thirring coupling, the vertex would either
diverge or vanish in the continuum limit. By tuning
a parameter in the regularization one can arrange that an apparent limit
is obtained as continuum is approached. Moreover, one can see whether
the correct numerical value is indeed obtained. 
(Actually, one uses the bare $e_0$ to
extract the dimensions and this might be the source of a systematic error.
If measurements turn out 
close enough to the exact number one hopes one can ignore this
error, at least at the present level of accuracy.)

Our numerical 
technique, the overlap, has been developed over several years starting
with [9] and being finalized in [10]. We added several new tricks to 
increase the efficiency of the simulation. 
In the overlap one computes two ground states
for each fermion multiplet, 
one corresponding to a positive mass $m_1$, 
and the other to a negative mass $m_2$. Previously [4,7,8] 
we took $m_1=-m_2$, but the positive mass can be made
infinite without spoiling the construction. At infinite mass
the needed ground state is known analytically and overall 
computation time gets reduced
by about one third. $m_2=-m$ is restricted to a bounded
range and shall be tuned (see later) to diminish the influence of 
induced Thirring couplings. 
To be sure, $m$ stays of the order of
$1/a$, where $a$ is the lattice spacing, so
the tuning really applies to $ma$. Another trick was already used in [4]
where the gauge fields were generated using a non-compact action 
(the fermions always see compact fields, so this is just a technicality). 
We create the instantons ``by hand'' and adjust the
measured vertex expectation value by the (analytically known)
probability for the gauge action to spontaneously generate an instanton.
Since instanton creation would be a Poisson process 
if left to the stochastic dynamics, a large statistical error
would be induced (measuring the probability to produce an instanton), so
the saving in computer time is substantial, by a factor of 10 roughly. 
A third trick
we employed was to generate the constant part of the vector potentials
by a weight equal to the fermion induced action in the continuum. 
This gave us a reduction by
a factor of about 3 in computer time. A fourth trick we used was to note
that charge 2 fermions ``need'' a finer lattice than charge 1 fermions. 
Thus, our
charge 2 fermions were living on a lattice of size $L\times L$ while
the charge 1 fermions were living on an embedded lattice of size $L/2
\times L/2$. (Both types of fermions interact with the same gauge field,
only one uses longer paths for parallel transport on the coarser lattice.)
Since there are four charge one fermions which have to be dealt with 
separately as they obey different boundary conditions, the time reduction is
large, however it comes at the expense of some systematics. A fifth 
trick we employed was related to the appearance of a derivative in the chiral vertex.
We extracted the analytical correction
coming from the finite difference representation of the derivative of $\eta$, 
and we scaled the measured numbers to compensate for this effect. 
Finally, to ameliorate possible $SU(4)$
breaking effects induced by the boundary conditions
we averaged over all index distributions
in the point-split, gauge invariant, lattice vertex.
We ended up using about one month of
computer time on a dedicated workstation
to produce the numerical results
below.

We started by measuring $<V_v >$ in the four flavor Schwinger
model. We found that picking $m=.5$ seemed to make Thirring effects
undetectable. Throughout the simulation we kept the physical size
of the torus fixed, adjusting the bare gauge coupling when the lattice size
is increased (i.e. when the lattice spacing is reduced). We chose 
${{e_0 l}\over\sqrt{\pi}}=1.5$ which corresponds to the volume size
in our simulation of the single flavor Schwinger model [8]
(all our tori had equal sides of size $l$). 
Analytical calculations in the continuum [11] show that 
finite size effects reduce $<V_v >$,
equal to $.00645$ in infinite volume, to $.0031$ at ${{e_0 l}\over\sqrt{\pi}}=1.5$. 

The basic method of measurement is the same as in [8], 
except diagonalization is now done by Householder's method rather
than Jacobi's. The
approach to the continuum limit can be seen from Table 1. Quite early,
a smooth monotonic behavior sets in which seems to asymptote to the
correct theoretical value within errors. Note that the sizes we quote are
for a lattice which has charge one fermions at each site, so one could
argue that these sizes should be doubled when comparing with
results in the chiral case. 

\def\tablerule{\noalign{\hrule}}

\def\htwo{height3pt&\omit && \omit &\cr}
\def\hthree{height4pt&\omit && \omit &\cr}
{\null\hfill 
\vbox{\offinterlineskip
\hrule
\halign{\vrule# & \hfil# & \vrule# & \hfil# & \vrule# \cr
\hthree
& \ \ $L$  && \ \ $<V_v>$  & \cr
\hthree
\tablerule\cr
\htwo
& \ \  10 && \ \ 0.0022(2) & \cr
\htwo
& \ \  12 && \ \ 0.0026(2) & \cr
\htwo
& \ \  14 && \ \ 0.0025(2)  & \cr
\htwo
& \ \  16 && \ \ 0.0026(2)  & \cr
\htwo
& \ \  18 && \ \ 0.0028(2) & \cr
\htwo
\hthree
\tablerule\cr
\hthree
& \ \  $\infty$ && \ \ 0.0031 & \cr
\htwo
\hthree}
\hrule}
\hfill}
\vskip .2cm
\centerline{
{\bf Table 1:} $ <V_v >$ for various lattice spacings (${l\over L}$) 
at $e_0 l /\sqrt{\pi} = 1.5$.} 
\centerline{ At $e_0 l /\sqrt{\pi} = \infty$ the continuum
value is .00645.}
\vskip .3cm

Our experience in the vector-like model 
led us to choose $m=.5$ also in the chiral simulation.
The infinite volume continuum value for the measured vertex in the
11112 model is .081, but a finite volume computation
in the continuum is not available yet. We can only guess that a 
similar reduction, by a factor of about ${{.00645}\over {.0031}}=2.1$ 
will occur in our finite system. 

Our results, corrected for point-split effects, are listed in Table 2. 
The non-negligible point-split correction factor also appears in Table 2. 
All in all the numbers we measure
seem to be very reasonable and compatible with what we know about the
exact result. We guess that the continuum limit of the vertex
in the chiral case,
taking into account the finite size corrections, is
${{.081}\over {2.1}}=.039$. For the lattice spacings that the chiral
model was simulated at, the vector model shows finite cutoff effects
that diminish the vertex value by 20\% - 30\%. This is entirely
compatible with our measurements in the chiral case. In summary,
the indication is that the overlap works quantitatively in the chiral 11112
model just as well as it does in the four flavor vector model.

We feel that our simulations provide sufficient 
evidence to justify the investment of resources 
needed to carry out a large scale simulation which would 
seek to work in significantly larger volumes and measure the spectrum 
directly. It is conceivable that further savings in computer time 
could be obtained by ``improving'' the lattice Hamiltonians used to
define the overlap.

\figure{1}{\captionone}{5.0}

\figure{2}{\captiontwo}{5.0}

\def\tablerule{\noalign{\hrule}}

\def\hfour{height3pt&\omit && \omit && \omit && \omit&\cr}
\def\hfive{height4pt&\omit && \omit && \omit && \omit &\cr}
{\null\hfill 
\vbox{\offinterlineskip
\hrule
\halign{\vrule#& \hfil# & \vrule# & \hfil# & 
\vrule# & \hfil# & \vrule# & \hfil# & \vrule# \cr
\hfive
& \ \ $L$  && \ \ $<V>_{Landau-gauge}$  && \ \ $<V>_{gauge-average} $  && \ \ point-split factor  & \cr
\hfive
\tablerule\cr
\hfive
& \ \  8 && \ \ 0.0251(06) && \ \ 0.0257(08) && \ \ 0.6610 & \cr
\hfour
& \ \  10 && \ \ 0.0256(07) && \ \ 0.0264(08) && \ \ 0.7397 & \cr
\hfour
& \ \  12 && \ \ 0.0288(11) && \ \ 0.0302(15) && \ \ 0.7940 & \cr
\hfour
& \ \  14 && \ \ 0.0305(13) && \ \ 0.0317(14) && \ \ 0.8328 & \cr
\hfour
& \ \  16 && \ \ 0.0316(12) && \ \ 0.0333(14) && \ \ 0.8615 & \cr
\hfour
\hfive}
\hrule}
\hfill}
\vskip .2cm
\centerline{
{\bf Table 2:} $ <V>$ for various lattice spacings (${l\over L}$) 
at $e_0 l /\sqrt{\pi} = 1.5$.}
\centerline { At $e_0 l /\sqrt{\pi} = \infty$ the continuum
value is .081.}
\vskip .3cm

Since one of the new ingredients in the chiral case
(and a very time consuming one) is the need to do gauge averaging
we show in Figures 1 and 2 two examples of the phase fluctuations
in the chiral observable, one at zero topology and the other in a
background carrying unit topological charge. Both examples are in 
statistically generic gauge field backgrounds.
We have explained in [4,10] that the gauge averaging should produce no
net phase and should have only an irrelevant (barring Thirring terms)
effect. 
In Table 2 we therefore show results both in a fixed
(Landau like) gauge (no gauge averaging) and including gauge averaging. 
There is no evidence that a discernible difference will be maintained 
in the continuum limit between the gauge fixed and gauge averaged
result. 
The phase distributions are quite feature-less, looking like 
periodic Gaussians, 
further indicating that indeed gauge invariance is restored without unwanted side effects (in [4] nontrivial topology was not
analyzed - we now see that gauge averaging works at least as well in the
topological charge one sector).

In addition to carrying out a large scale simulation of the two models
there are some less demanding directions for future work: One should
complete the solution of the Abelian model by finding exact expressions for
't Hooft's vertex on a finite torus and also include the effects of
a Thirring term. In the same context it might be interesting to compute
explicitly the form factors of the massless composites, as seen,
for example, by a small, slowly varying external gauge field coupling to
their $SU(4)$ charges. In the non-Abelian model one should examine further
the possibility to obtain some exact results about its massless content.
Also, there should exist some simple finite size dependent measurements that
could count efficiently the number of massless particle types in either
model.

\vskip .2cm

\noindent {\bf Acknowledgments}

This research was supported in part by the DOE under grants 
\# DE-FG06-91ER40614 (RN),
\# DE-FG06-90ER40561 (RN) and 
\# DE-FG05-96ER40559 (HN). One of us (R.N.) would like to thank D.B. Kaplan 
for discussions.

\vskip 1cm

\noindent{\bf References}
\vskip .2cm
\item{[1]} G. 't Hooft, Lectures at Cargese Summer Institute (Sept. 1979).
\item{[2]} S. Dimopoulos, S. Raby, L. Susskind, Nucl. Phys. B173 (1980) 208; 
S. Raby, S. Dimopoulos, L. Susskind, Nucl. Phys. B169 (1980) 373.
\item{[3]} T. Banks, Y. Frishman, S. Yankielowicz, Nucl. Phys. B191 (1981) 493.
\item{[4]}  R. Narayanan, H. Neuberger, Nucl. Phys. B, in print, 
hep-lat \# 9607081.
\item{[5]} For a more detailed review see section 10 in [10].
\item{[6]} S. Coleman, ``The Uses of Instantons'', Erice Subnucl. 1977:805.
\item{[7]} R. Narayanan, H. Neuberger, P. Vranas, Nucl. Phys. B 
(Proc. Suppl.) 47 (1996) 596.
\item{[8]} R. Narayanan, H. Neuberger, P. Vranas, Phys. Lett. B353 (1995) 507.
\item{[9]}  R. Narayanan, H. Neuberger, Phys. Lett. B302 (1993) 62.
\item{[10]}  R. Narayanan, H. Neuberger, Nucl. Phys. B443 (1995) 305. 
\item{[11]} This requires a trivial generalization to several flavors 
of the work in I. Sachs, A. Wipf, Helv. Phys. Acta 65 (1992) 652. 
\vfill
\eject

\end